\documentclass[11pt]{article}
\usepackage{jheppub}

\usepackage[space]{grffile}
\usepackage{pdfpages}

\usepackage{slashed,physics}
\usepackage{graphicx}
\usepackage{amsmath,amssymb,graphicx}
\usepackage{textcomp} 
\usepackage{gensymb} 
\usepackage{epsf,color}
\usepackage[dvipsnames]{xcolor}
\usepackage{hyperref}
\usepackage{cancel}
\usepackage{framed}
\usepackage{hyperref}
\usepackage{float}
\usepackage{multirow}
\usepackage{subfigure}
\definecolor{nicered}{rgb}{0.5,0.,0.}
\definecolor{nicegreen}{rgb}{0.,0.5,0.}
\definecolor{niceblue}{rgb}{0.,0.,0.5}
\hypersetup{colorlinks,citecolor=nicegreen,linkcolor=nicered,urlcolor=niceblue}
\numberwithin{equation}{section}
\newcommand{\beq}{\begin{equation}}
\newcommand{\eeq}{\end{equation}}
\newcommand{\bea}{\begin{eqnarray}}
\newcommand{\eea}{\end{eqnarray}}
\newcommand{\bear}{\begin{eqnarray}}
\newcommand{\eear}{\end{eqnarray}}

\newcommand{\ba}{\begin{array}}
\newcommand{\ea}{\end{array}}

\title{Higgs decay to charmonia and the charm-quark Yukawa coupling}

\author[a]{Tao Han}
\author[a]{Adam K. Leibovich}
\author[a,b]{Yang Ma}
\author[a,c]{Xiao-Ze Tan}

\affiliation[a]{Pittsburgh Particle Physics, Astrophysics and Cosmology Center,\\
Department of Physics and Astronomy, University of Pittsburgh, \\Pittsburgh, PA 15260, USA}

\affiliation[b]{INFN, Sezione di Bologna, via Irnerio 46, 40126 Bologna, Italy}

\affiliation[c]{Department of Physics and Center for Field Theory and Particle Physics, \\Fudan University, Shanghai, 200438, China}

\emailAdd{than@pitt.edu}
\emailAdd{akl2@pitt.edu}
\emailAdd{yang.ma@bo.infn.it}
\emailAdd{xz\_tan@fudan.edu.cn}

\preprint{PITT-PACC 2215}

\abstract{
  After the great triumph of the Higgs discovery in 2012, the next target at the energy frontier will be to study the Higgs properties and to search for the next scale beyond the SM. Experimentally, the $H\to c \bar{c}$ channel would be extremely difficult to dig out because of both the weak Yukawa coupling and the daunting SM di-jet background. We propose to test the charm-quark Yukawa coupling at the LHC and future hadron colliders with the Higgs boson decay to $J/\psi$ via the charm-quark fragmentation. Using the non-relativistic quantum chromodynamics (NRQCD), we study the charmonia production via the Higgs boson decay channel $ H \to c \  \bar{c} + J/\psi $(or $ \eta_c $), where both the color-singlet and color-octet contributions are considered. Our result opens another door to improve determinations at the LHC of the Higgs Yukawa couplings: the final state from this decay mode is quite distinctive with $J/\psi\to e^+e^-,\, \mu^+\mu^-$ and the branching fraction is enhanced by the charm-quark fragmentation mechanism.
  }

\makeatletter
\gdef\@fpheader{}
\makeatother
\begin{document}
\maketitle

\section{Introduction}
With the discovery of the Higgs boson at the CERN Large Hadron Collider (LHC) in 2012 \cite{Aad:2012tfa,Chatrchyan:2012ufa}, the particle spectrum of the Standard Model (SM) is confirmed. 
As Higgs is believed to be the portal to new physics beyond the Standard Model (BSM), its interactions to the other SM particles need to be measured as precisely as possible.
While the measurements show that the Yukawa couplings of the Higgs to the third generation fermions ($t{\bar t}$, $b{\bar b}$, and $\tau {\bar \tau}$) are consistent with the SM prediction very well 
, the value of the Higgs-charm coupling still remains to be confirmed.

The current measurements on the charm quark Yukawa coupling $y_c$ given by ATLAS and CMS are based on the $p p \rightarrow V H$ channel, and the constraints on the $\kappa_c =y_c/y_c^{\rm SM}$ are $|\kappa_c|<8.5$ \cite{ATLAS:2022ers} and $1.1<|\kappa_c|<5.5$ \cite{CMS:2022psv}. 
In this paper, we suggest that it is possible to measure $y_c$ at the High-Luminosity LHC (HL-LHC) via the Higgs decays $H \to c + {\bar c} + J/\psi$ based on the recent calculation in Ref.~\cite{Han:2022rwq}.

\section{Higgs decay to charmonia in the Standard Model}
The calculation of charmonia production via the Higgs decay can be done within the non-relativistic quantum chromodynamics (NRQCD) framework \cite{Bodwin:1994jh}, and the decay width of Higgs boson into quarkonium can be written as
\begin{eqnarray}
  \Gamma =\sum_\mathbb{N}  {\hat \Gamma}_\mathbb{N}(H \to(Q {\bar Q})[n]+X)\times \langle {\cal O}^h[\mathbb{N}] \rangle,
  \label{eq:Gamma}
\end{eqnarray}
where $\mathbb{N}$ stands for the involved $Q{\bar Q}$ Fock state with quantum numbers $n(^{2S+1}L_J^{\rm [color]})$. 
While the short-distance coefficient (SDC) ${\hat \Gamma}_\mathbb{N}$ can be perturbatively calculated following the Feynman diagrams, the long-distance matrix elements (LDMEs) $\langle {\cal O}^h[\mathbb{N}] \rangle$ need to be extracted from either the wave function at origin or the experimental data. 
For $J/\psi$ production, the contributions from both the color-singlet state ($^3S_1^{[1]}$) and the color-octet ones ($^3S_1^{[8]}$, $^1S_0^{[8]}$, and $^3P_J^{[8]}$) are considered.

The idea of taking $J/\psi \to \mu^+\mu^-$ as an effective trigger for measuring $y_c$ was firstly suggested in Ref.~\cite{Bodwin:2013gca,Bodwin:2014bpa}. However, the branching fraction of the suggested decay channel ($H\to J/\psi + \gamma$) is too small, and the dominance of the ``vector meson dominance'' (VMD)  $\gamma^*\to J/\psi$ makes the above process rather insensitive to the $y_c$. 

\begin{figure}[htb]
  \centering
  \includegraphics[width=.24\textwidth]{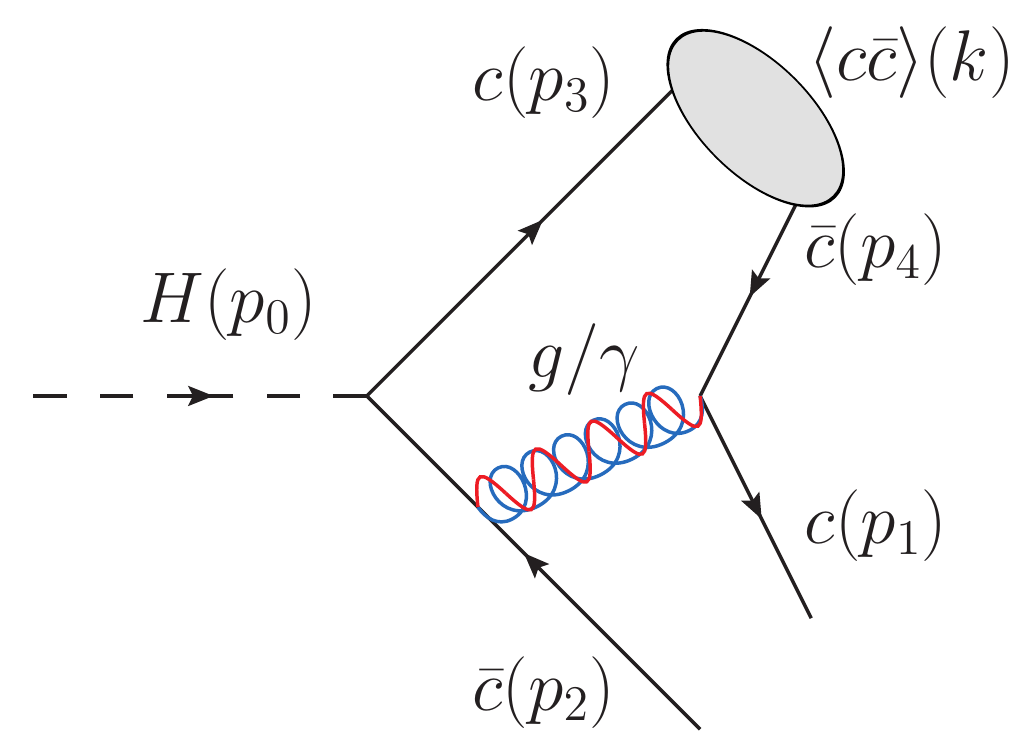}
  \includegraphics[width=.24\textwidth]{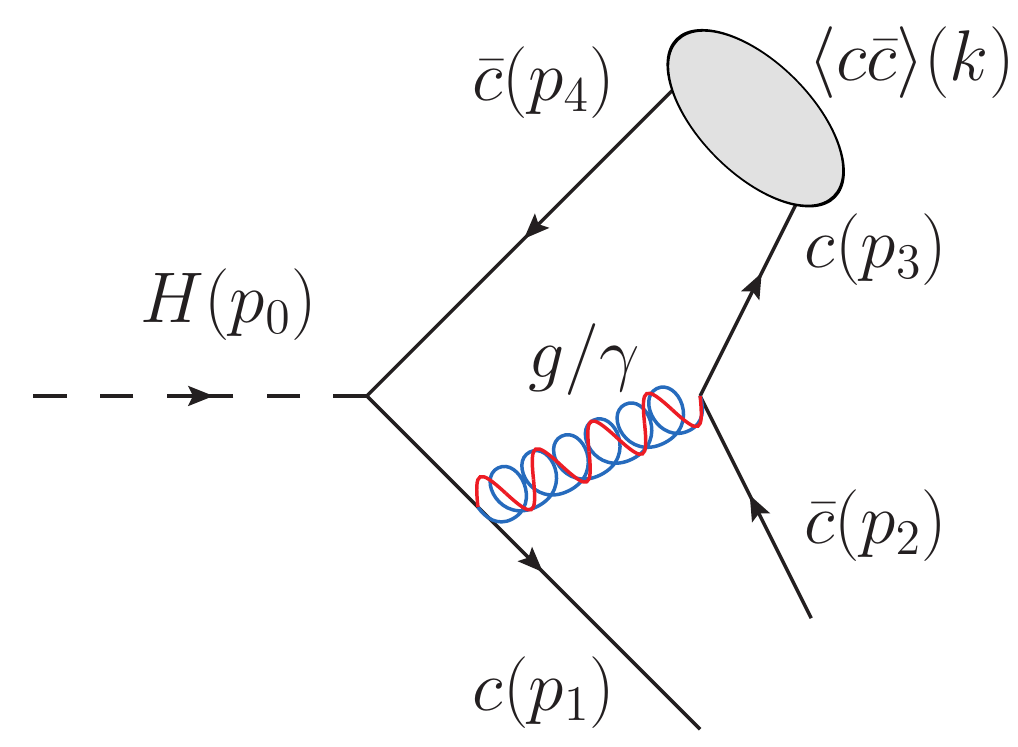}
  \includegraphics[width=.24\textwidth]{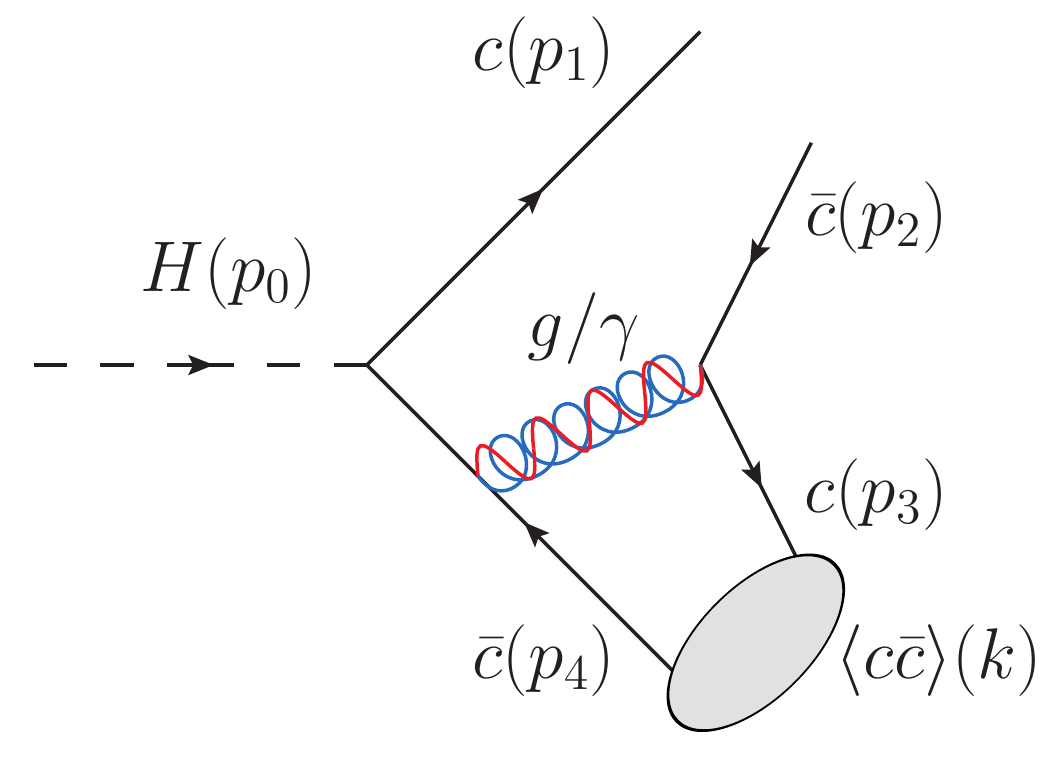}
  \includegraphics[width=.24\textwidth]{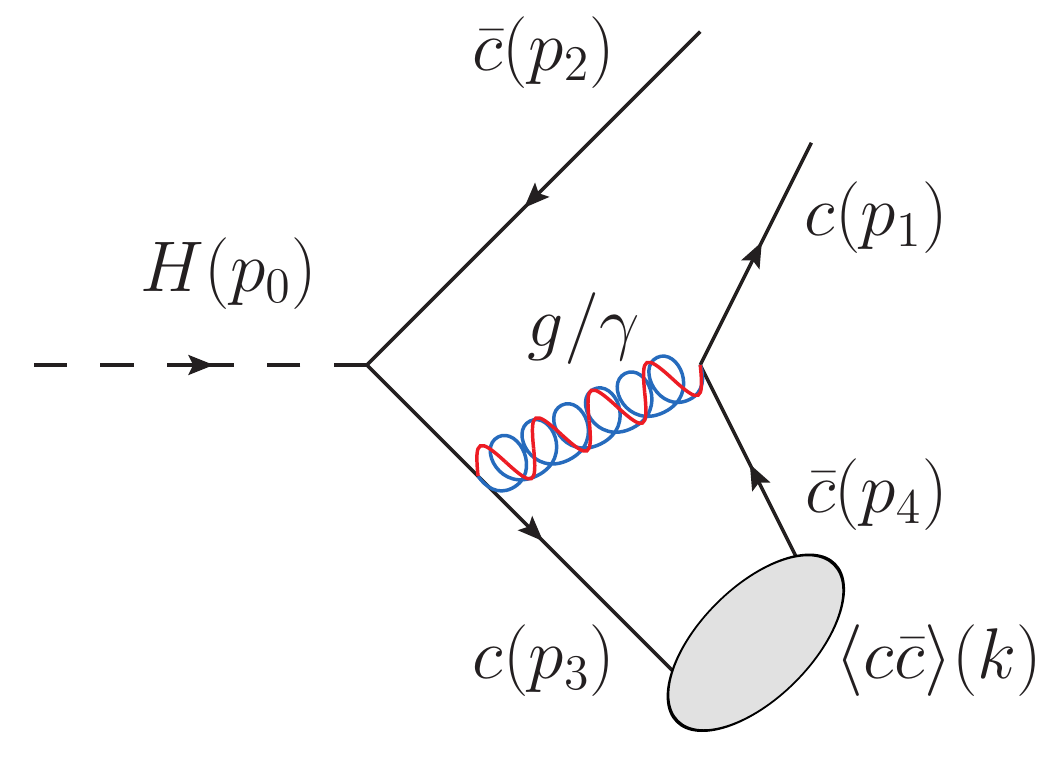}
  \includegraphics[width=.25\textwidth]{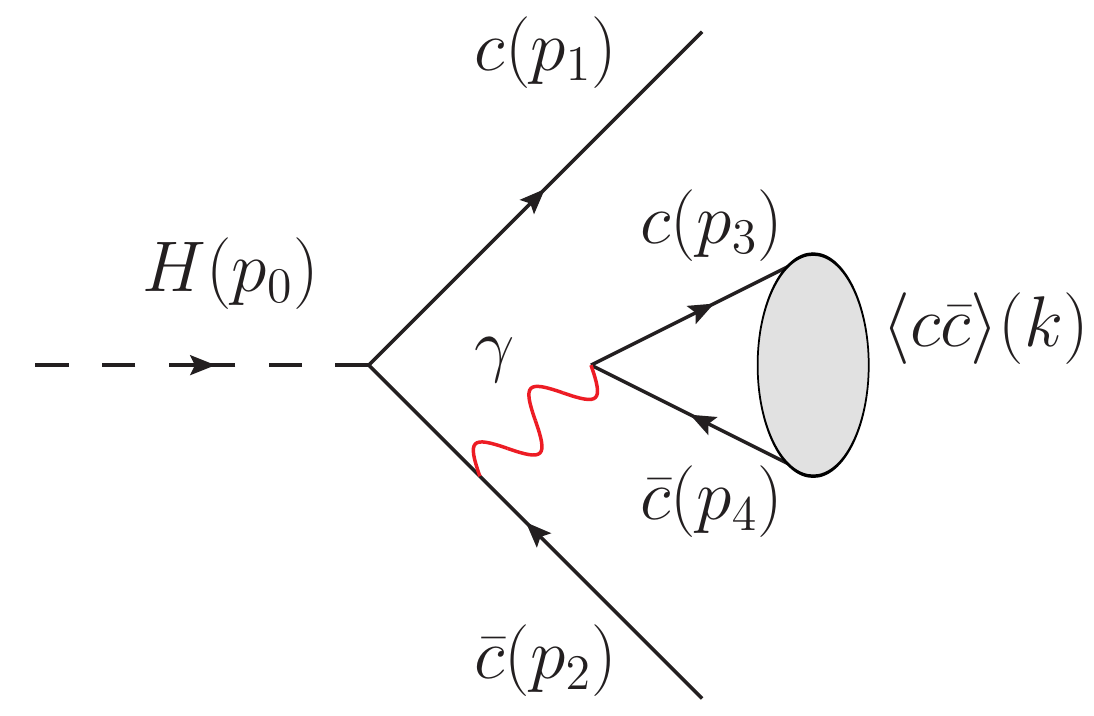}
  \includegraphics[width=.25\textwidth]{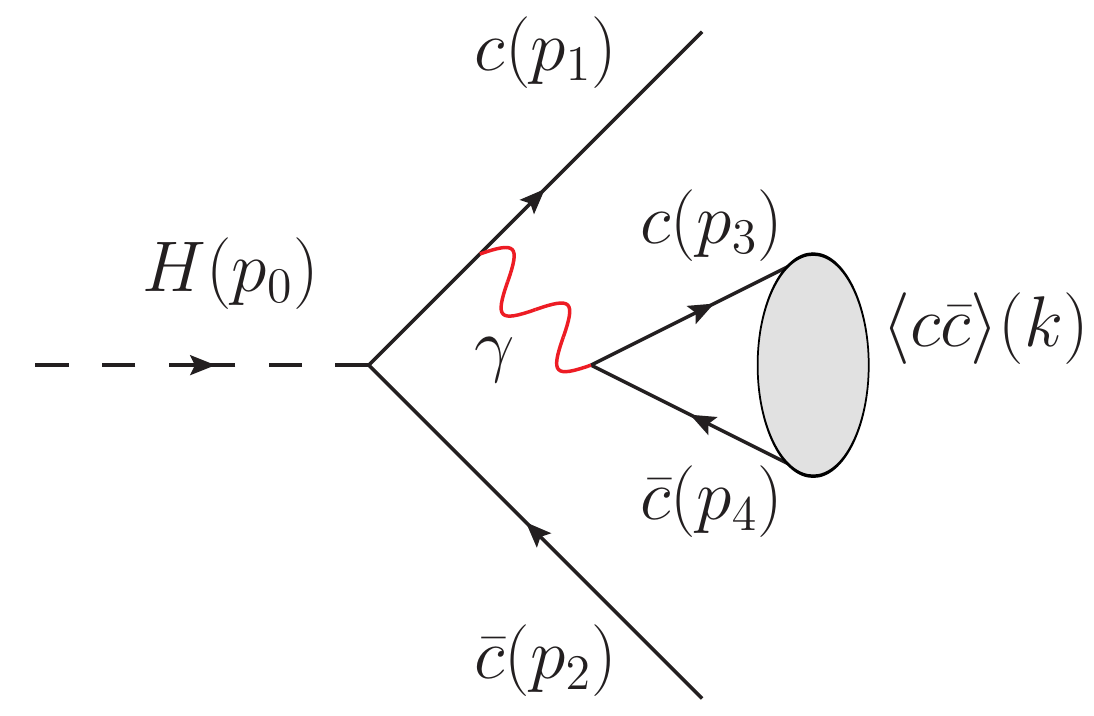}
  \includegraphics[width=.22\textwidth]{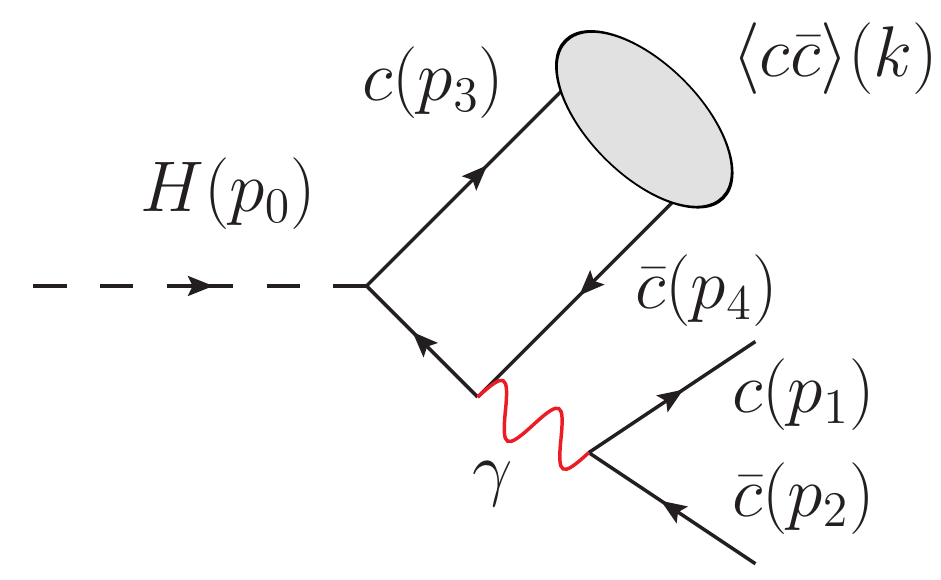}
  \includegraphics[width=.22\textwidth]{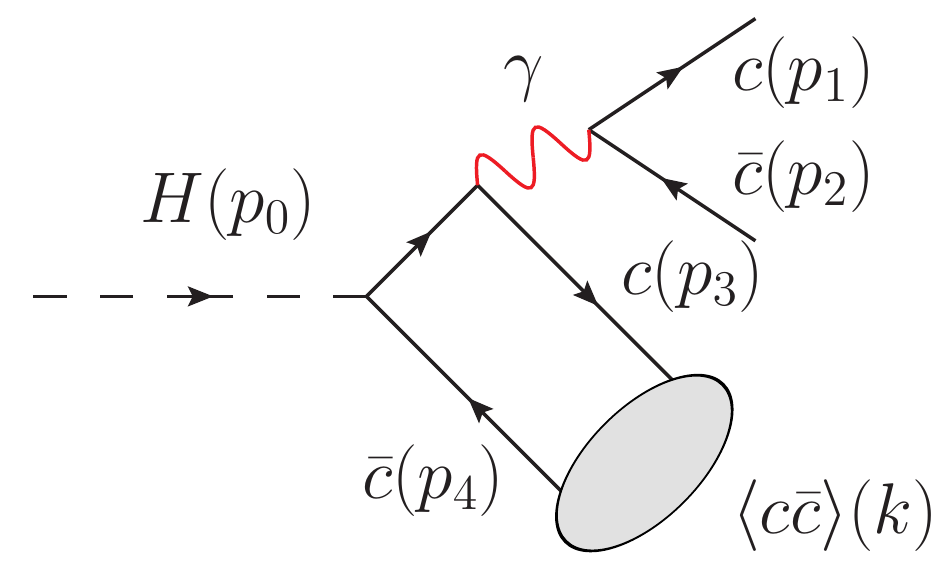}
  \includegraphics[width=.25\textwidth]{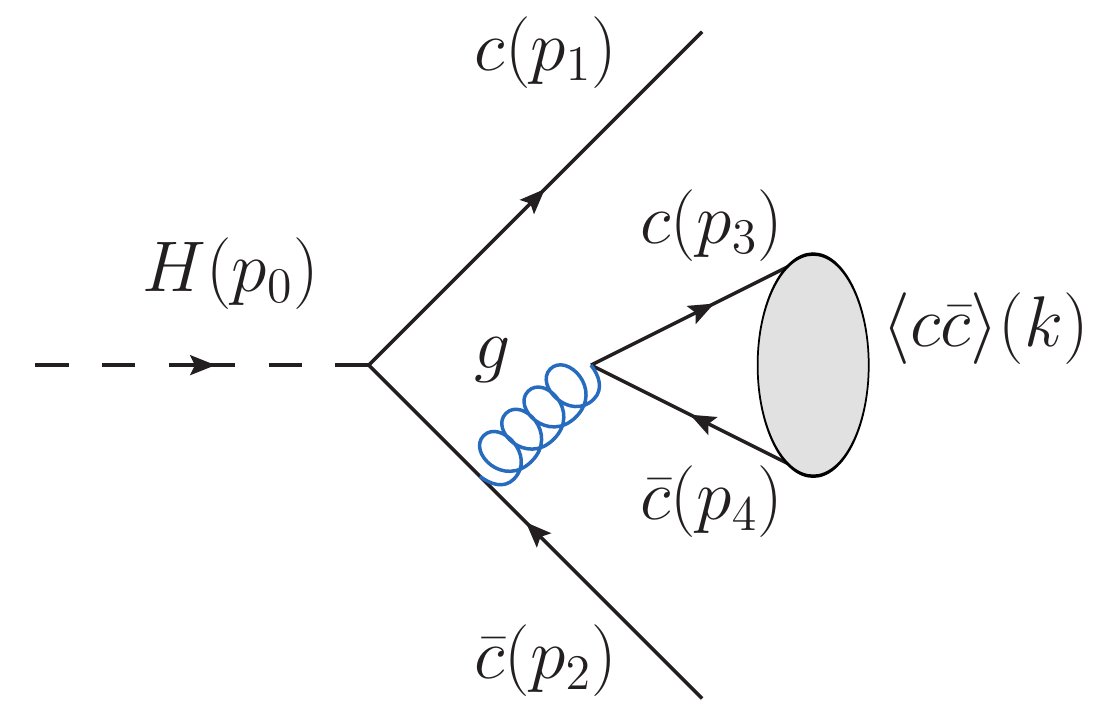}
  \includegraphics[width=.25\textwidth]{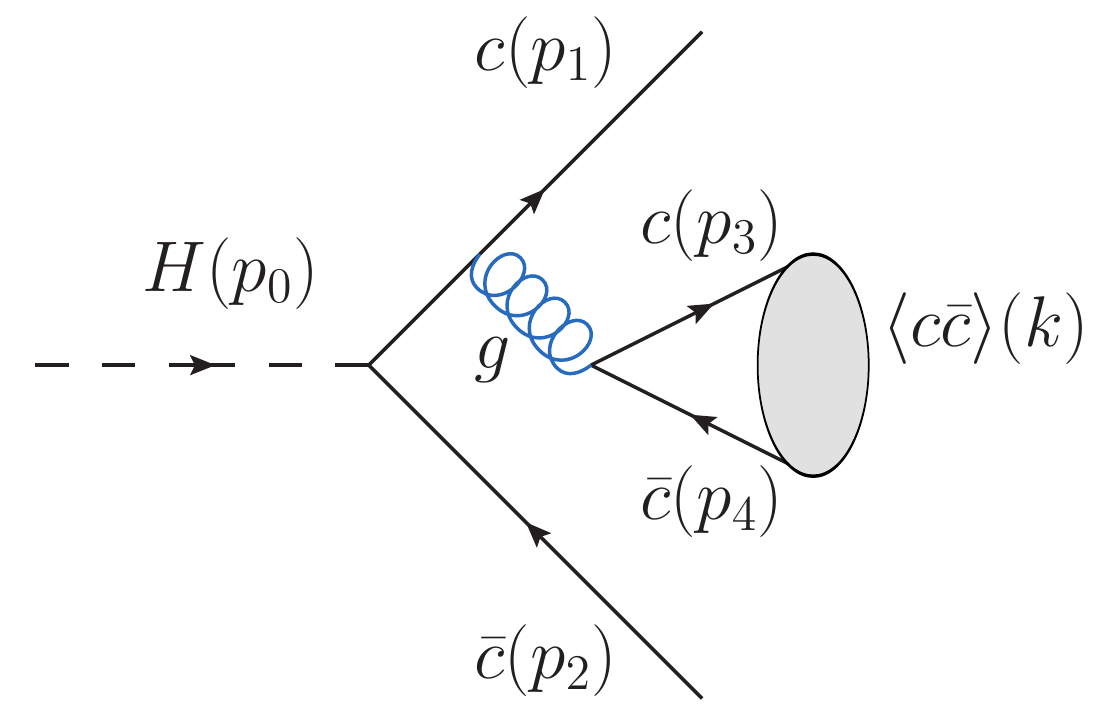}
  \includegraphics[width=.22\textwidth]{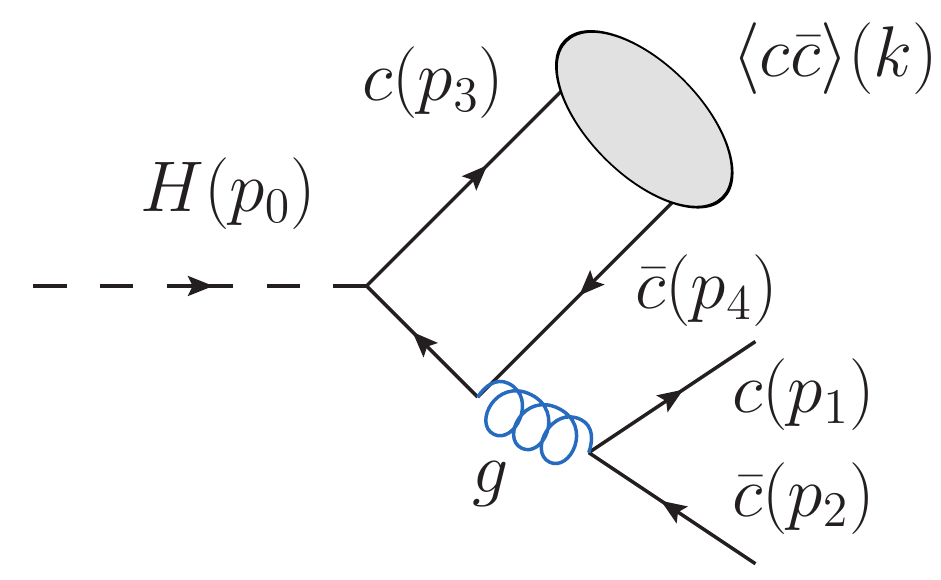}
  \includegraphics[width=.22\textwidth]{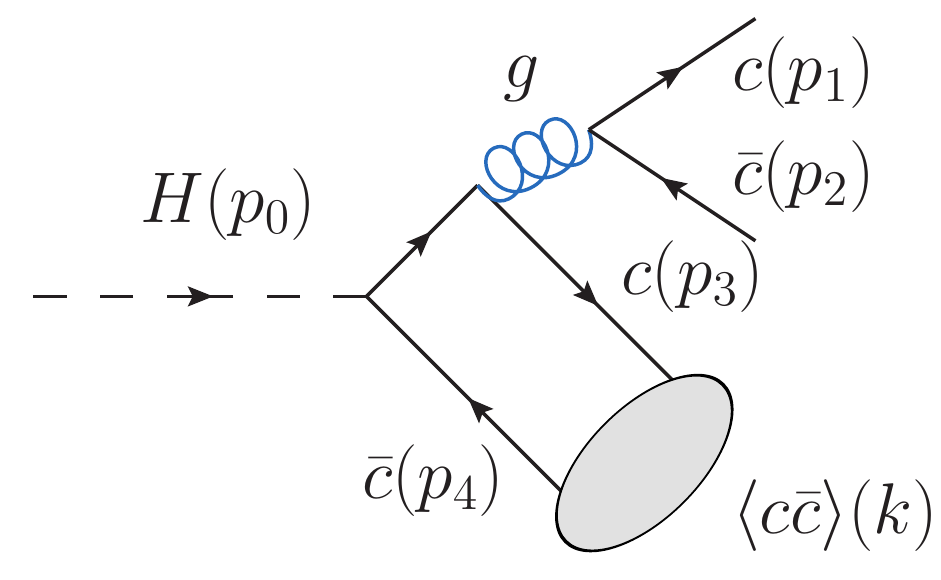}
  \caption{Typical Feynman diagrams for $H \to c {\bar c} + J/\psi$ at leading-order. The first row is the charm quark fragmentation mechanism, the first two diagrams in the second row are for the single photon fragmentation, and the first two diagrams in the third row are for the single gluon fragmentation. }
  \label{fig:Frag}
\end{figure}

\begin{figure}[htb]
  \centering
  \includegraphics[width=.24\textwidth]{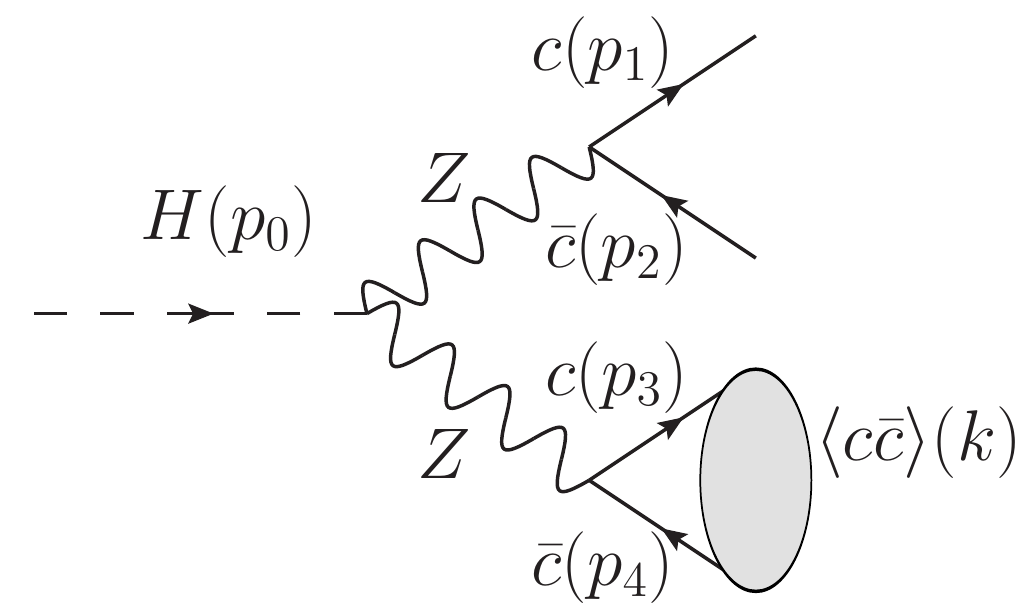}
  \includegraphics[width=.24\textwidth]{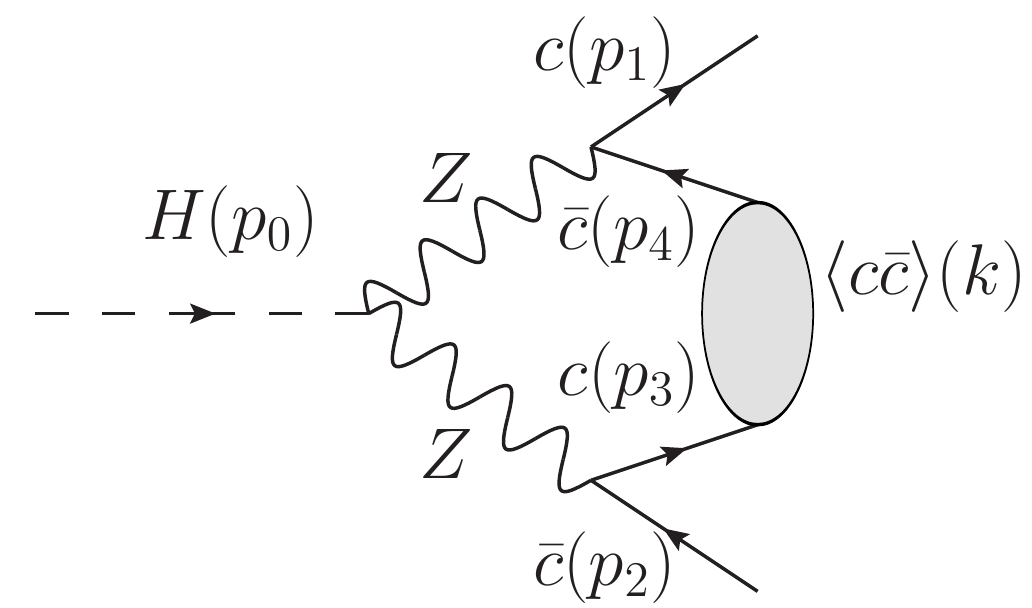}
  \caption{Feynman diagrams for $H \to c {\bar c} + J/\psi$ through the $HZZ$ vertex.}
  \label{fig:HZZ}
\end{figure}

\begin{figure}[tb]
  \centering
  \includegraphics[width=.48\textwidth]{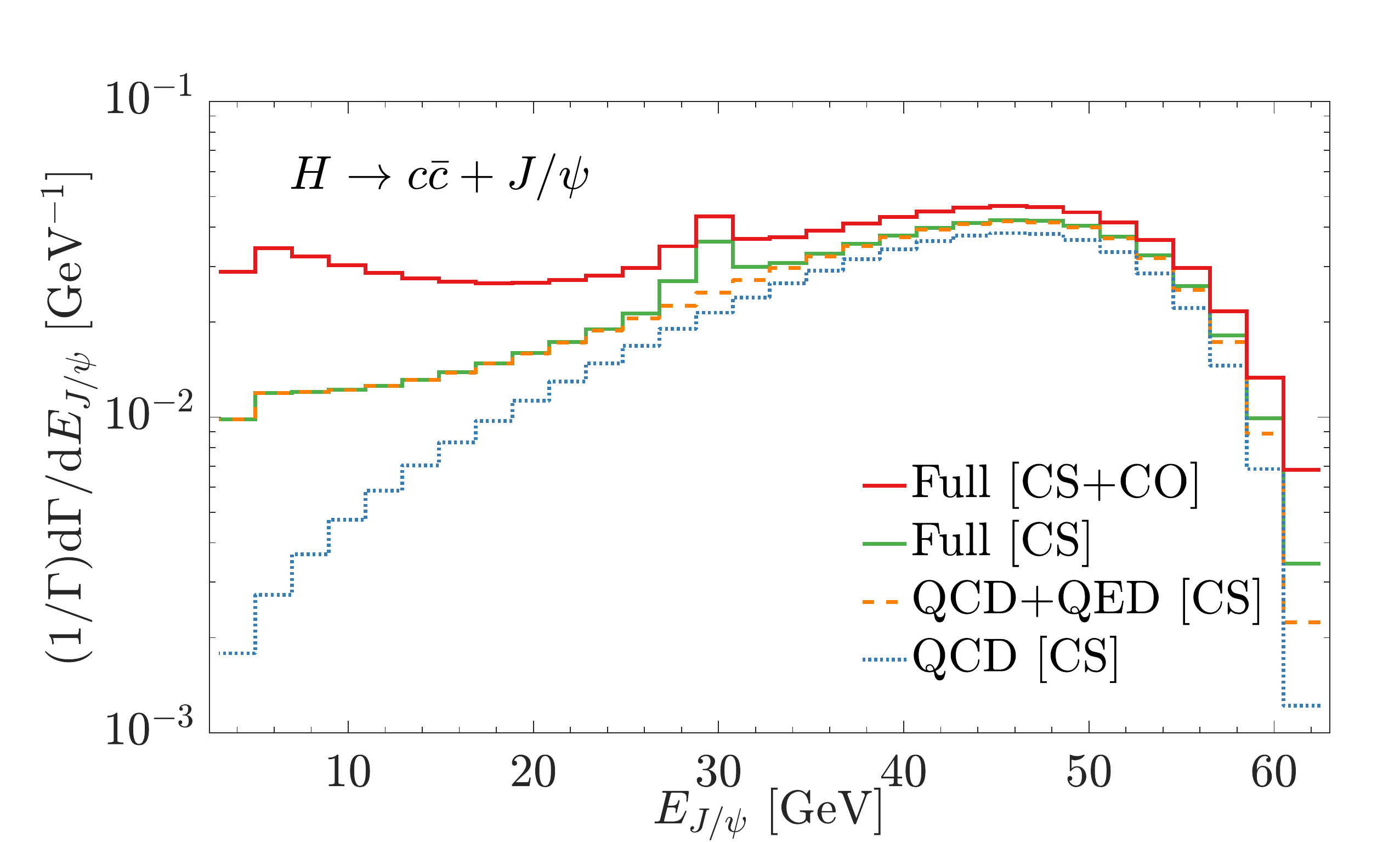}
  \includegraphics[width=.48\textwidth]{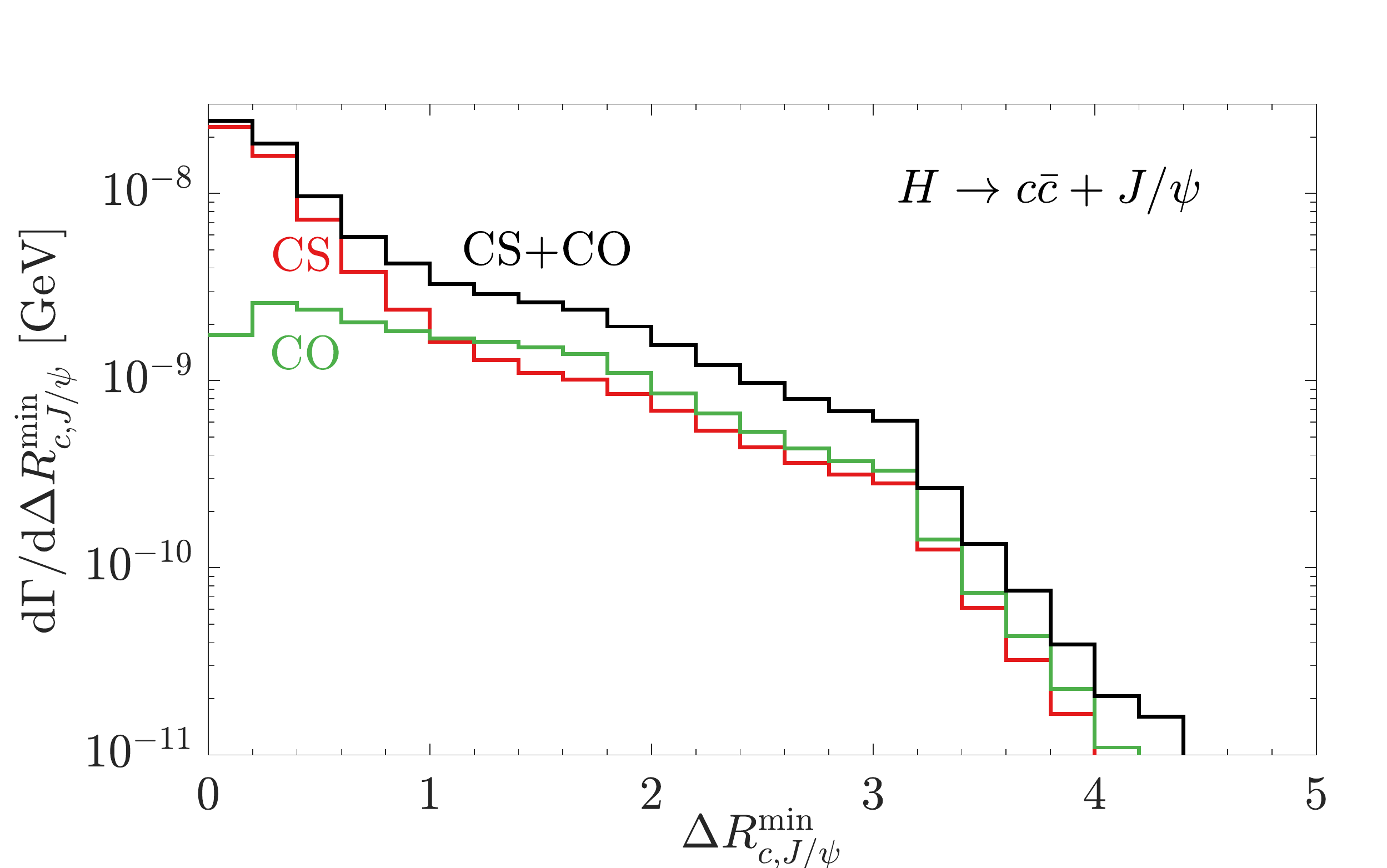}
  \caption{(a) The $J/\psi$ energy distributions, where blue dotted and orange dashed curves are for color-singlet (CS) QCD only and QCD+QED contributions, the red (green) solid curve is for the sum of both the CS and the color-octet (CO) contributions (full CS contribution). All curves are normalized using the full leading order decay width in Table \ref{tab:SMBodwin}. (b) The smaller angular distance between $J/\psi$ and one of the free charm quark ${\rm d}\Gamma/{\rm d}\Delta R_{c,J/\psi}^{\rm min}$, where the black, red, and green solid curves are for the full leading order contribution, the CS only, and CO only, respectively.} 
  \label{fig:Edis_decom}
\end{figure}

Another potentially promising decay channel is $H \to c {\bar c} + J/\psi$.
The dominant contribution to this decay process is the fragmentation mechanism built upon the $H\to c{\bar c}$ decay, and there are power/logarithmic enhancements due to the fragmentations of the $c$ quark, the photon, and the gluon splittings \cite{Han:2022rwq}. 
The typical Feynman diagrams for $H \to c {\bar c} + J/\psi$ are listed out in Fig.~\ref{fig:Frag}.
At leading order, there is also the contribution from the electroweak sector via the $HZZ$ vertex whose the Feynman diagrams are shown in Fig.~\ref{fig:HZZ}.
Take all the above mechanisms into account, and implement the numerical values for the SM parameters, $\alpha_s(Q)$ running, and $m_c(Q)$ running using \texttt{para} \cite{Han:2020uid,para}, we have the decay widths and the corresponding branching fractions in Table~\ref{tab:SMBodwin}\footnote{The amplitude square can be analytically simplified using \texttt{FeynCalc} \cite{Mertig:1990an,Shtabovenko:2016sxi,Shtabovenko:2020gxv}.}.
The $J/\psi$ energy distribution ${\rm d}\Gamma/{\rm d} E_{J/\psi}$ is presented in Fig.~\ref{fig:Edis_decom}(a), where it is shown that the photon/gluon fragmentation diagrams have dramatic enhancements on $^3S_1^{[1]}$ and $^3S_1^{[8]}$ in the low $J/\psi$ energy region, and the charm quark fragmentation dominates in the relative high energy region. The $Z$ resonance peak due to Fig.~\ref{fig:HZZ} (a) appears  at $E_{J/\psi} = {\frac{1}{2}} m_H (1-m_Z^2/m_H^2 + 4m_c^2/m_H^2)\approx 30$ GeV. From observation point of view, it is also interesting to show the smaller angular distance between $J/\psi$ and one of the free charm quark ${\rm d}\Gamma/{\rm d}\Delta R_{c,J/\psi}^{\rm min}$ in Fig.~\ref{fig:Edis_decom}(b). 

\begin{table}[htb]
	\caption{The decomposed numerical values of $\Gamma (H \to c  {\bar c} + J/\psi)$ and the corresponding branching fractions. The color-singlet and color-octet contributions are labeled as CS and CO, respectively.}
	\label{tab:SMBodwin}
	\center
	\scalebox{0.9}{
		\begin{tabular}{lccccc}
			\hline
			\multicolumn{1}{c}{}                   & QCD [CS]              & QCD+QED [CS]          & Full [CS]             & Full [CO]             & Full [CS+CO]          \\ \hline
			$\Gamma(H\to c{\bar c}+J/\psi)$ (GeV) & $4.8\times 10^{-8}$ & $5.8\times 10^{-8}$ & $6.1\times 10^{-8}$ & $2.2\times 10^{-8}$ & $8.3\times 10^{-8}$ \\
			${\rm BR}(H\to c{\bar c}+J/\psi)$     & $1.2\times 10^{-5}$ & $1.4\times 10^{-5}$ & $1.5\times 10^{-5}$ & $5.3\times 10^{-6}$ & $2.0\times 10^{-5}$ \\
			\hline
		\end{tabular}
	}
\end{table}

\section{Probe the charm quark Yukawa coupling}
\begin{figure}[tb]
  \centering
  \includegraphics[width=.48\textwidth]{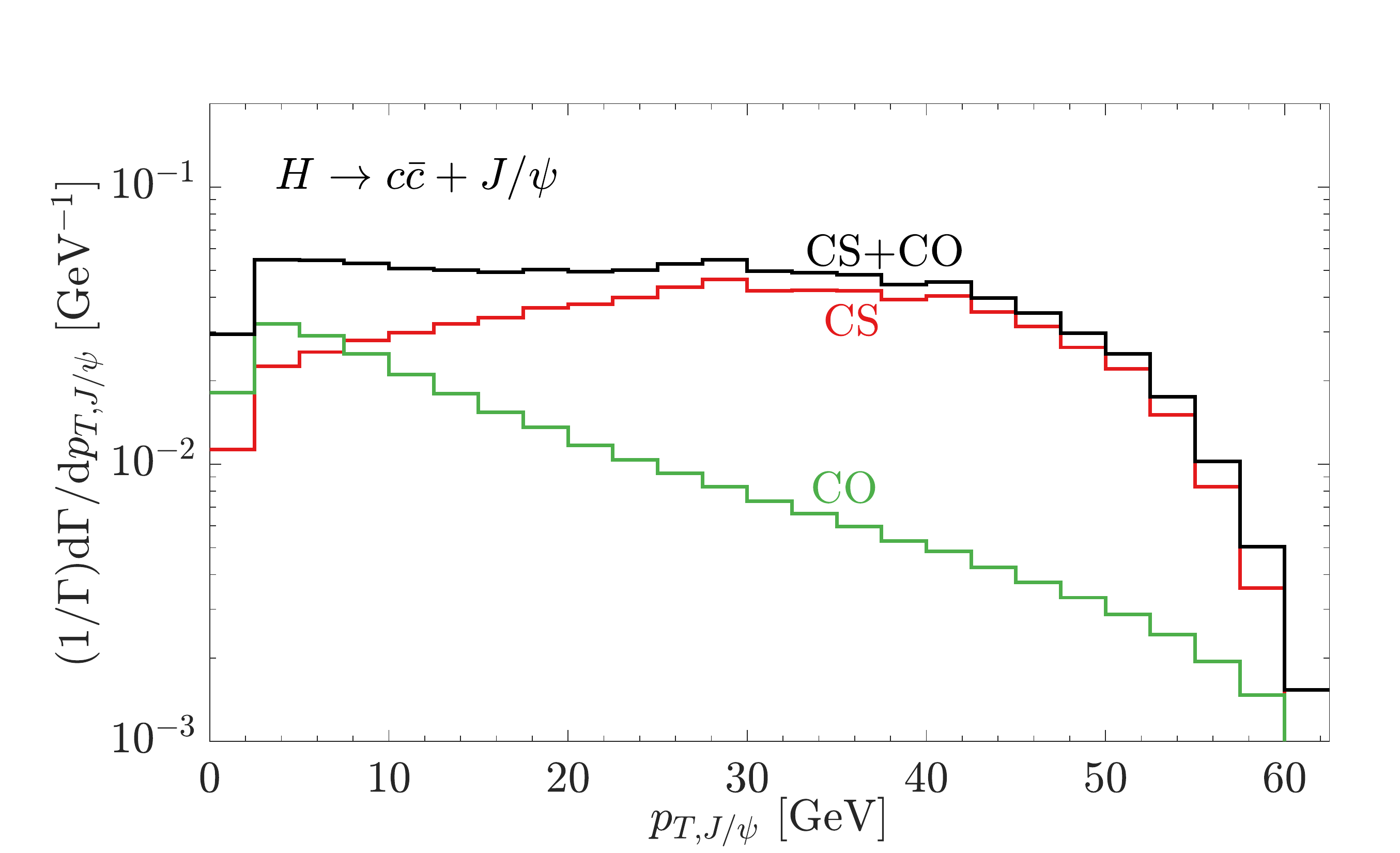}
  \includegraphics[width=.48\textwidth]{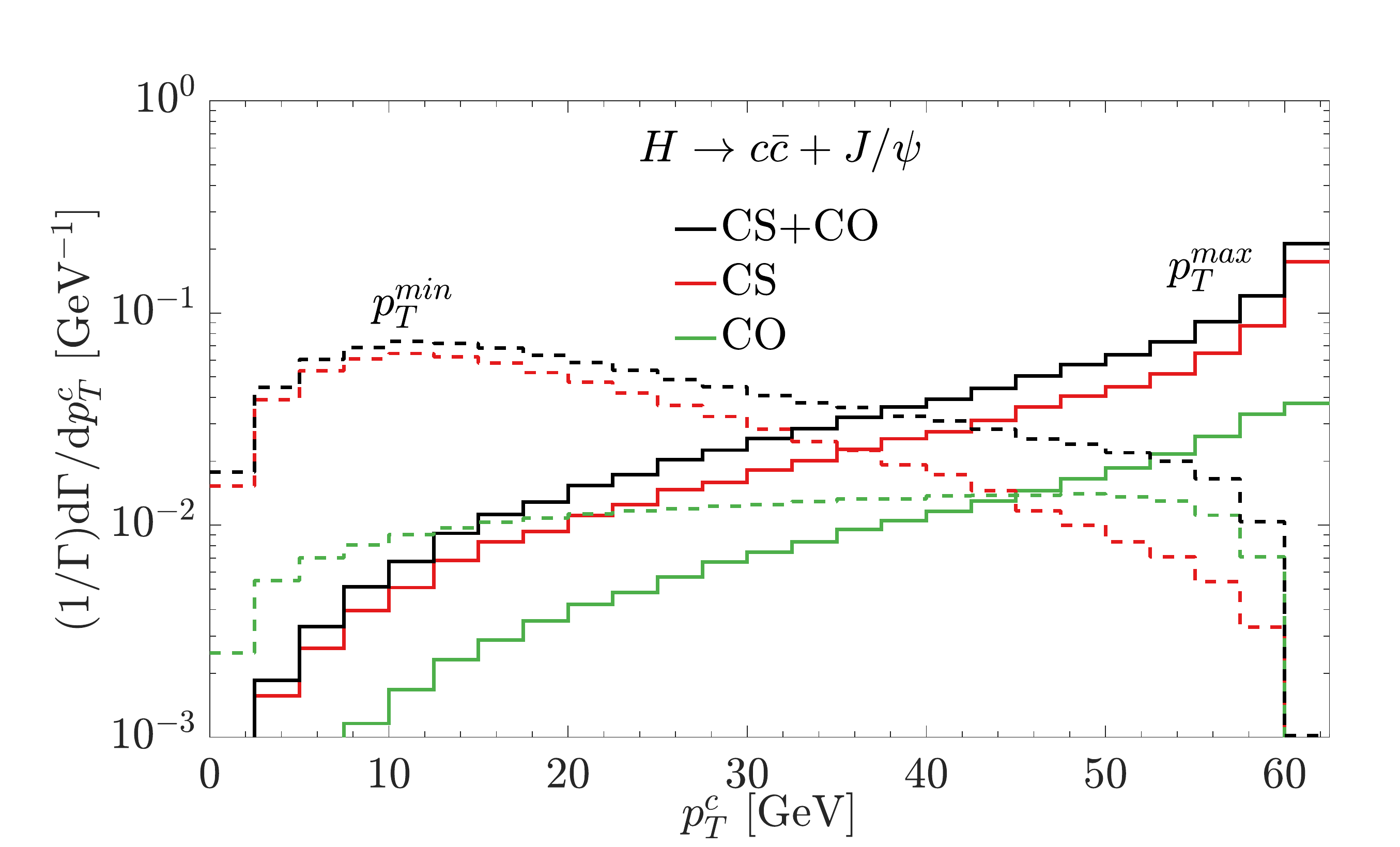}
  \caption{Transverse momentum distributions in the Higgs rest frame $H\to c\bar c+ J/\psi$: 
(a) and (b) are for $J/\psi$ distribution and the free charm quark distributions associated with $J/\psi$, respectively, where the solid curves are for the $p_T^{max}$ and dashed curves are for the $p_T^{min}$ distribution. The red, green and black curves are for CS, CO, and the full leading order result. All curves are normalized using the full leading order decay width  in Table \ref{tab:SMBodwin}.}
  \label{fig:ptdis}
\end{figure}

According to the numerical results in Ref. \cite{Han:2022rwq}, it is proper to employ the $\kappa$ framework and parameterize
\begin{eqnarray}
  y_c=\kappa_c y_c^{\rm SM},~~~ {\rm BR} \approx \kappa_c^2\ {\rm BR}^{\rm SM}
\end{eqnarray} 
For the expected $L\sim 3\, {\rm ab}^{-1}$ HL-LHC \cite{Apollinari:2017cqg}, consider the Higgs production cross section $ \sigma_H^{}\approx 50$ pb, one could write the anticipated number of events as\footnote{The 12\% branching fraction for $J/\psi \to \mu^+\mu^-, e^+e^-$ has been included.} 
\begin{equation}
  N = L \sigma_H^{}\ \epsilon\ {\rm BR}(c\bar c+\ell^+\ell^-) \approx 12\ \kappa_c^2 \times {\frac{L}{{\rm ab}^{-1}}}\times {\frac{\epsilon}{10\%}}, 
  \end{equation}
where we have introduced the efficiency $\epsilon$ to detect the final state $c\bar c+\ell^+\ell^-\ (\ell=\mu,e)$.
Based on the $p_T$ distributions in Fig.~\ref{fig:ptdis}, we can roughly estimate the kinematic acceptance of $50\%$.
If we also consider the double charm tagging of $(40\%)^2$, we then can assume $\epsilon \sim 10\%$.
Take the statistical error only $\delta N \sim \sqrt{N}$, we have
\begin{equation}
  \Delta\kappa_c \approx 15\% \times ({\frac{L}{{\rm ab}^{-1}}}\times {\frac{\epsilon}{10\%}})^{-1/2}.  \label{eq:Deltakappac}
\end{equation}

\begin{figure}[tb]
\centering
\includegraphics[width=.24\textwidth]{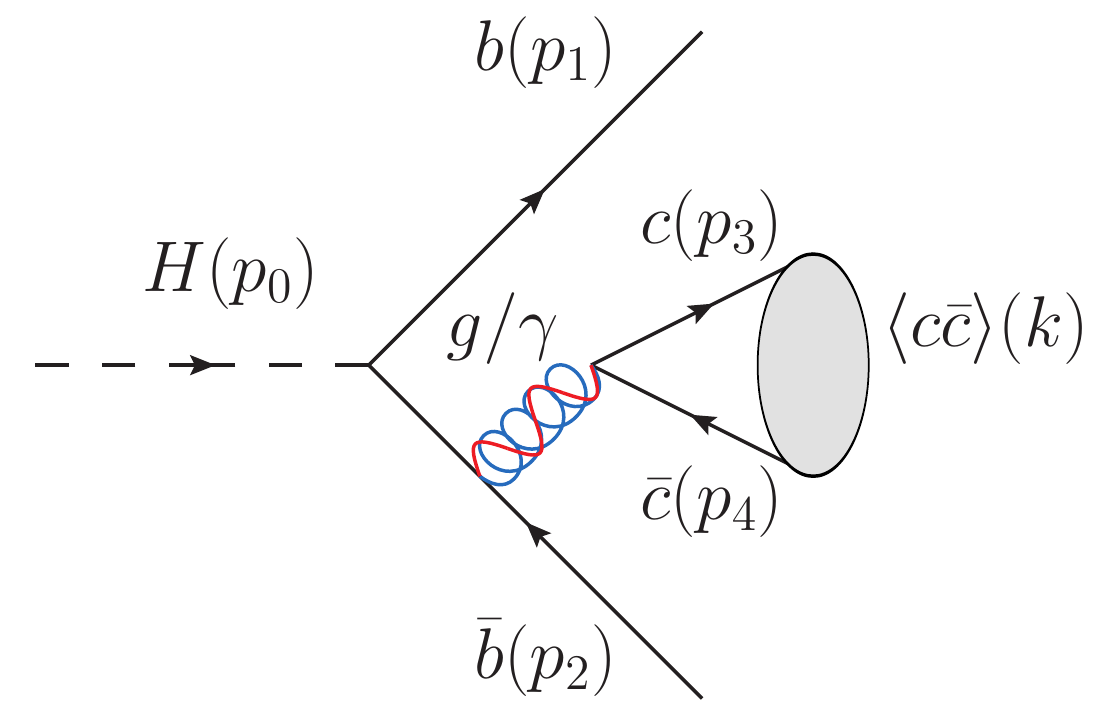}
\includegraphics[width=.24\textwidth]{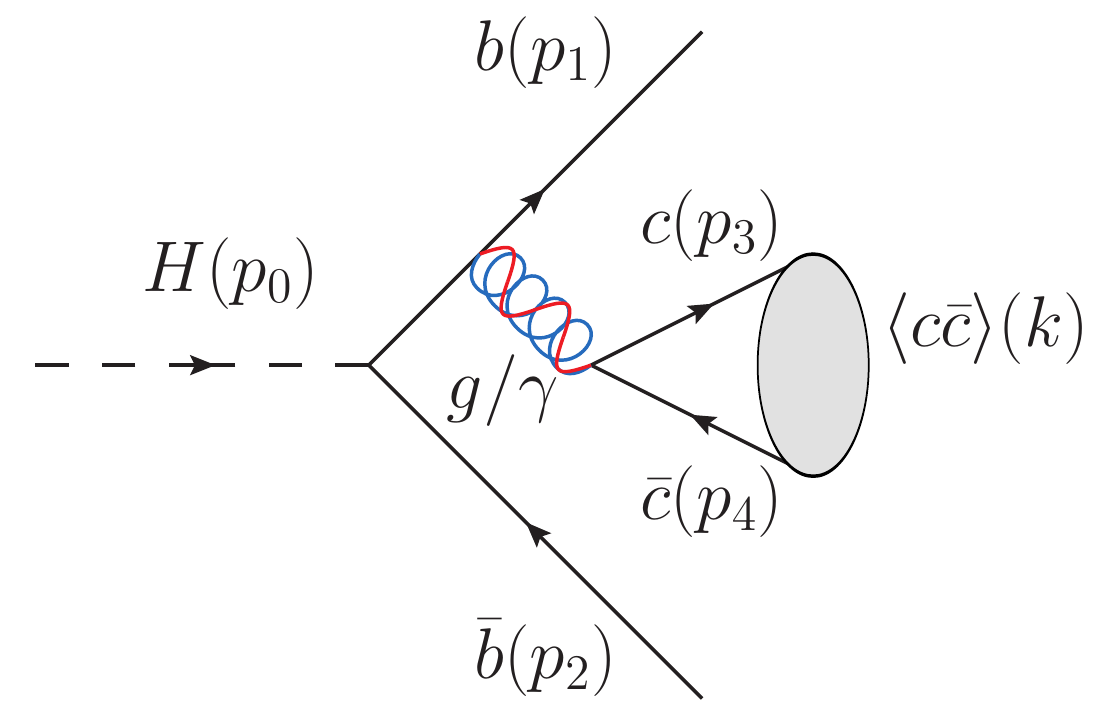}
\includegraphics[width=.24\textwidth]{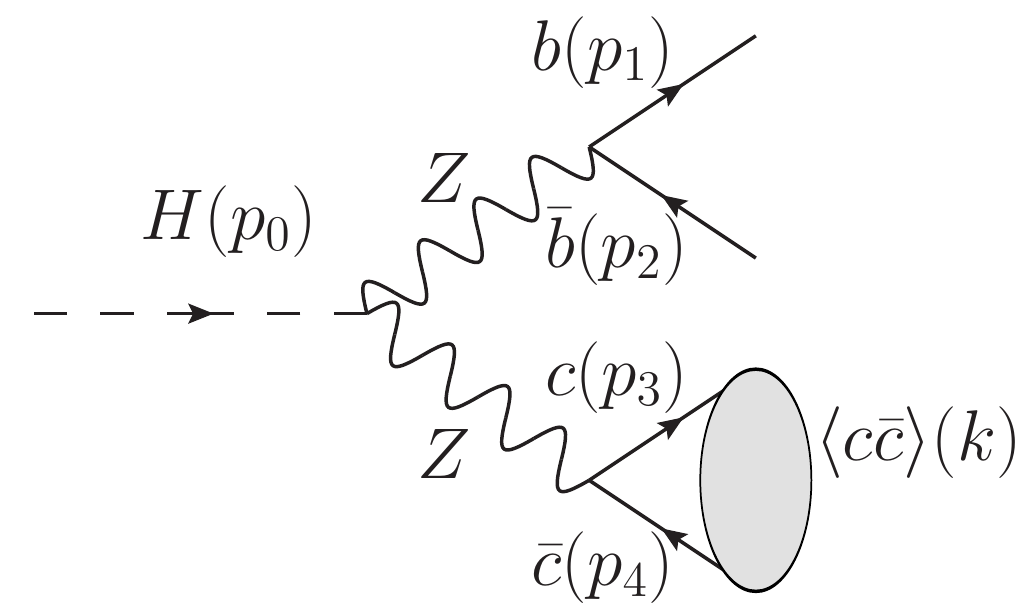}
\includegraphics[width=.24\textwidth]{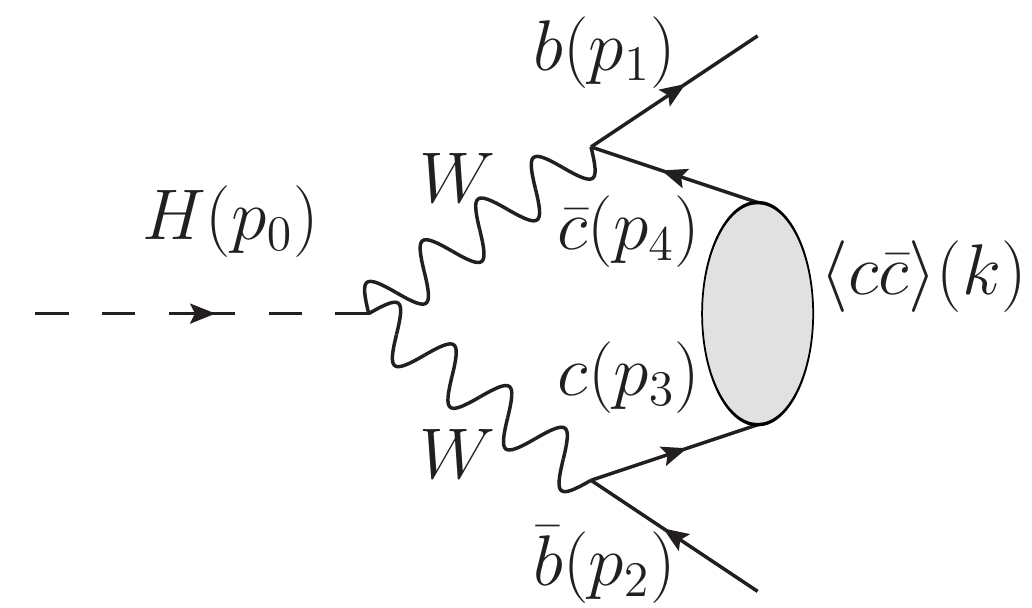}
  \caption{Feynman diagrams for $ H\to b\bar{b}+ J/\psi (\eta_c)$ production. The gluon diagrams in (a, b) only contribute to $ ^3 S_1 ^{[8] }$, while the photon ones only contribute to $ ^3 S_1 ^{[1]} $. (c) is nonzero only for the CS states.}\label{fig:Hbb}
\end{figure}

For realistic experimental searches at the LHC, there exist large backgrounds. 
The prompt $J/\psi$ production at the LHC has been measured to be ${\rm BR}(J/\psi \to \mu^+\mu^-)\times\sigma(pp\to J/\psi)\simeq 860~$pb for $20\leq p_T\leq 150~{\rm GeV}$, with a data sample of $2.3~{\rm fb}^{-1}$ by CMS \cite{CMS:2017dju}.
Employing double charm tagging would likely reduce this background.
The leading irreducible background signal is from the QCD production of $J/\psi$ plus charm jets, whose cross section falls dramatically versus the transverse momentum dropping by $4$ orders of magnitude at $p_T\simeq 20\,{\rm GeV}$ \cite{Artoisenet:2007xi}. We expect to reduce this background using some suitable kinematic cuts.
Another background is from the $H\to b{\bar b} + J/\psi$ decay due to the relatively large bottom Yukawa coupling, as shown in Fig.~\ref{fig:Hbb}. The decay width is calculated to be
\begin{eqnarray}
  {\rm BR}(H\to b{\bar b}+J/\psi)= 8.6 \times 10^{-5},
\end{eqnarray}
which is around 4 times larger than that of $c\bar c + J/\psi$. An effective charm-tagging need to be applied to separate these two in the future analysis. 
For a rough estimate, assume there are $10,000$ background events after the proper selection cuts at the HL-LHC, one could reach a $2\sigma$ sensitivity for the coupling $\kappa_c\approx 2.4$. 
Detailed analyses for the detector and systematic effects are left for a future work.

\section{Summary}
After the Higgs discovery in 2012, the measurements on Higgs properties has become one of the most important tasks in particle physics. Though the Yukawa couplings of the Higgs to the third-generation fermions are measured to be consistent with the SM prediction, the charm quark Yukawa coupling $y_c$ remains to be determined.
We propose to measure $y_c$ using the Higgs decay process $H\to c{\bar c} + J/\psi$, and have calculated the branching fraction
\begin{eqnarray}
	{\rm BR}(H\to c{\bar c}+J/\psi)\approx 2.0 \times 10^{-5}.
\end{eqnarray}
Considering possible background signals and kinematic cuts, it is possible to reach a $2\sigma$ sensitivity for the coupling $\kappa_c \approx 2.4$ at the $L\sim 3\,{\rm ab}^{-1}$ HL-LHC. Detailed studies including the detector and systematic effects would be called for to reach a quantitative conclusion.

\acknowledgments
This work was supported in part by the U.S.~Department of Energy under grant No.~DE-SC0007914, 
U.S.~National Science Foundation under Grant No.~PHY-2112829, and in part by the PITT PACC. 
The support provided by China Scholarship Council (CSC) during the visit of Xiao-Ze Tan to PITT PACC is also acknowledged.

\bibliographystyle{JHEP}
\bibliography{ref.bib}

\end{document}